\documentstyle[twoside,fleqn,espcrc2]{article}
\newcommand{\gt} {>}
\newcommand{\A} {{\cal A}}
\newcommand{\B} {{\cal B}}

\newcommand{\AmS}{{\protect\the\textfont2
  A\kern-.1667em\lower.5ex\hbox{M}\kern-.125emS}}

\title{First order phase transitions of spin systems\thanks{Talk given
at the XII International Symposium on Lattice Field Theory,
LATTICE '94, Bielefeld Germany, September 27 - October 1 st 1994.
\vskip .1cm Preprint SPhT 94-164
\vskip .1cm hep-lat/9501003}}

\author{Alain BILLOIRE\address{Service de Physique Th\'eorique
de Saclay,\\
        91191 Gif sur Yvette France}}

\begin{document}

\begin{abstract}

I review  some numerical ways to determine the parameters of
systems close to a first order phase transition point: energy and
specific heat of the coexisting phases and interface tension.
Numerical examples are given for the 2-d $q$ states
Potts model.
\end{abstract}

\maketitle

\section{INTRODUCTION}

First order phase transitions are easily described in terms of the
competition between two phases (see e.g. \cite{Binder_review}).
Consider for definiteness a temperature driven
phase transition between a disordered
phase and a q-fold degenerate ordered phase of free energy
$f_d(\beta)$ and $f_o(\beta)$.
In  order to use numerical methods, one needs to know the
behavior of
statistical averages in a finite volume, close to the transition point.
This was more or less correctly guessed by several authors
\cite{Bind1,Bind2}, and put on rigorous footing (at least for a class
of models including the large $q$ Potts model) in Ref
\cite{Borgs_deux,Borgs_Miracle} where
the partition function is shown to behave  like

\begin{eqnarray}
\label{Z}
 Z(\beta,L)  &=&   e^{- L^d \beta f_d(\beta)} + q e^{- L^d \beta
f_o(\beta)}\\
\nonumber &+&   {\cal O}(e^{-b L})  e^{- \beta f(\beta) L^d }
\hskip .2cm ;\  b > 0
\end{eqnarray}

This formula is correct for periodic boundary conditions,
$f_d(\beta)$ and $f_o(\beta)$ are $L$ independent functions, and
$f(\beta) = min \{ f_d(\beta), f_o(\beta) \}$. Although there is no
analytical continuation of the free energy of one phase into the
other
phase, both $f_d(\beta)$ and $f_o(\beta)$ are shown to be several
times differentiable at $\beta_t$, the (infinite volume) transition
point where $f_o=f_d$. As a consequence the finite size
behavior of any moment of the energy is obtained by
differentiating Eq. \ref{Z}.  For example, the maximum of the
energy fluctuation behaves like

\begin{equation}
\label{max}
(<E^2> - <E>^2)_{max} = \A + \B L^{-d} + {\cal O} (L^{-2d})
\end{equation}

with similar results for $<S^2> - <S>^2$,
$BL={1\over 3}(1-{<E^4> \over <E^2>^2})$ (the Binder Landau
cumulant),
$1-{<E^2> \over <E>^2}$ \dots
The constants $\A$ and $\B$ in Eq.\ref{max} are
simple functions of $E_o$, $E_d$, $C_o$ and $C_d$,
the (infinite volume limit) energy and specific heat of the pure
phases at $\beta_t$.
Identical results for the first two terms in
the $1/L^d$ expansion can be obtained using the Binder Landau
ansatz\cite{Bind2} for the energy probability density $P_L(E)$.

\begin{eqnarray}
\label{P(E)}
P_L(E)     & = & N \Biggl[ q {a_o \over \sqrt{C_o}} \displaystyle
  \  e^{\ \displaystyle-{{ L^d \beta_t^2 (E-E_o-C_o\delta
T)^2}\over{2C_o}}}\\
\nonumber  & + & \ \ \ \   {a_d \over \sqrt{C_d}} \displaystyle
\    e^{\ \displaystyle-{{ L^d \beta_t^2 (E-E_d-C_d\delta
T)^2}\over{2C_d}}}
\Biggr] \\
\nonumber   \delta T &= & {1\over\beta}-{1\over\beta_t}
\end{eqnarray}

which can be obtained from Eq.\ref{Z} by inverse Laplace
transform,
neglecting the ${\cal O} (e^{-bL})$ terms.
In order to estimate $\A$ and $\B$ from numerical data,
the presence of $1/L^{2d }$ corrections in Eq.\ref{max} is a
nuisance. As  first
noticed in \cite{Borgs_Miracle}, averages estimated at $\beta_t$ are
better
behaved, for example the energy fluctuation

\begin{eqnarray}
(<E^2> - <E>^2)_{max} &=& \tilde\A + \tilde\B L^{-d} \\
\nonumber &+&  {\cal O} (e^{-bL})
\hskip .2cm ;\  b > 0
\label{betat}
\end{eqnarray}

The energy itself behaves like,

\begin{equation}
<E>_L(\beta_t) = <E>_{\infty}(\beta_t) + {\cal O}(e^{-b L})
\hskip .2cm ;\  b > 0
\end{equation}

\begin{figure}[htbp]            
\vskip 7cm
\includegraphics{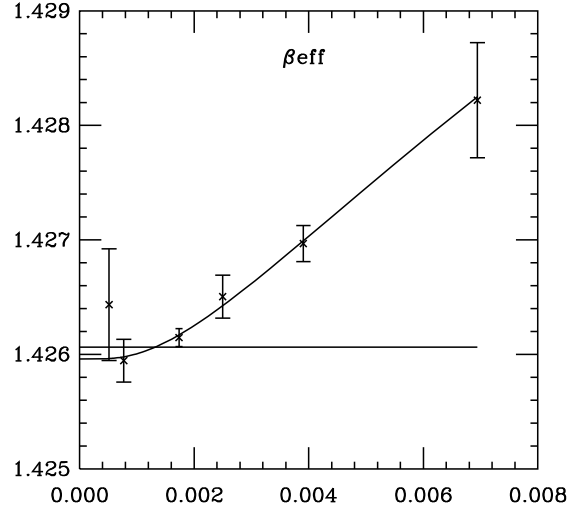}
\caption{Estimate of the infinite volume transition temperature
of the $q=10$ Potts model
obtained from the equality of  the internal energy of the
following pairs of lattices:
$\{ 12,16 \}$,$\{ 16,20\}$,$\{20,24\}$,$\{
24,36\}$,$\{36,44\}$,$\{44,50\}$,
as function of $1/L^2$, where $L$ is the smallest value of each
pair.
The curve drawn is a fit to the shape $a+b e^{-c L}$. Note that the
(unfounded) shape $a + b/L^2$ would give a good fit of the data
but an incorrect estimate for $\beta_t$.}
\label{Fig.beta_eff}
\end{figure}

A simple  powerful method to determine the value for
$\beta_t$ follows \cite{Borgs_Miracle}: perform simulations on
increasing
sized lattices, at a value close to $\beta_t$. Extrapolate
to neighboring values of $\beta$ using the Swendsen Wang
reweighting method\cite{Ferrenberg_Swendsen}.
$\beta_t$  is determined as the fixed point of the set of curves
$\{ <E>_L(\beta) \}$. This is
similar to Binder's cumulant method\cite{Binder_cumulant} to
determine a second order phase
transition point using the fourth order cumulant:
$1-{<S^4> \over 3<S^2>^2}$. However in the current case, the
estimate
has only exponentially small corrections.
Application\cite{Paper_q10} of
this method to the 2-d $q=10$ Potts model can be found in
Fig.\ref{Fig.beta_eff}. This is a strong first order phase transition
(see
Tab.\ref{table1}), the estimated value for $\beta_t$ is $1.42596
(7)$ in
excellent agreement with the exact value $\ln(\sqrt{10}+1) \sim
1.42606$.

\begin{figure}[htbp]            
\vskip 7cm
\includegraphics{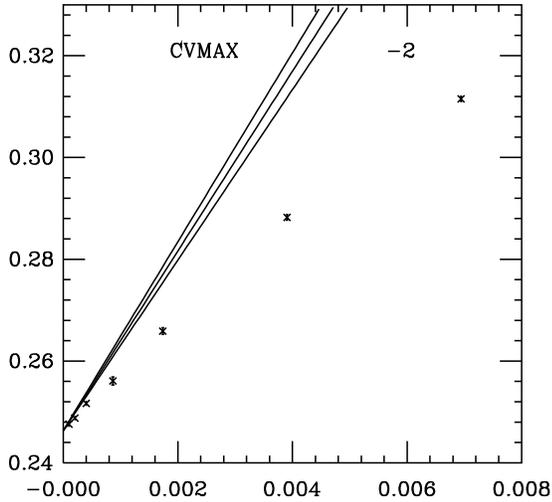}
\caption{The maximum of the  specific heat divided by $L^2$,
as a function of $1/L^2$, for the $q=10$ 2-d Potts model. The three
curves are drawn using the  values $C_o = 18.06 \times
\{.95,1.00,1.05\}$.}
\label{Fig.CV.max}
\end{figure}

\begin{figure}[htbp]            
\vskip 7cm
\includegraphics{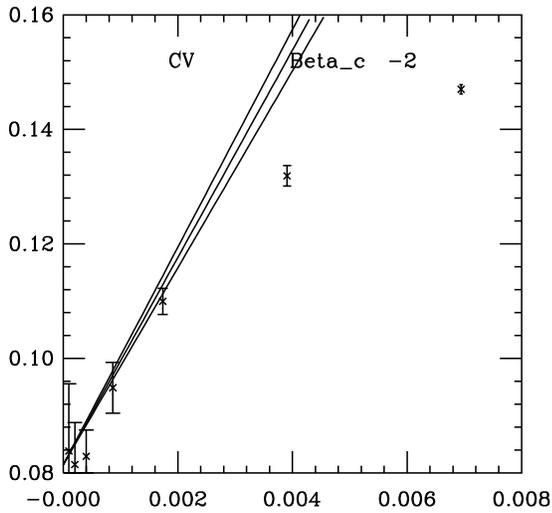}
\caption{$CV(\beta_t)/L^2$, as a function of $1/L^2$,
for the $q=10$ 2-d Potts model. The three
curves are drawn using the  values $C_o = 18.06 \times
\{.95,1.00,1.05\}$.}
\label{Fig.CV.beta}
\end{figure}

\begin{table} [htp]
\begin{small}
\begin{center}\begin{tabular}{|lrrrr|}\hline
q  &$C_o$& $E_d-E_o$\cite{Baxter}   & $C_d-C_o$  & $\xi_{\beta_t^-}$\cite{BW}\\
[3pt]\hline
 5  &                 & 0.05292  & 0.0326    &  2512        \\
 7  &$71.3  (10)$     & 0.35328 & 0.2234    &  48.09       \\
10  &$18.06  (4)$     & 0.69605 & .44763    &  10.56  	   \\
20  &$5.362  (3)$     & 1.19416 & .77139    &   2.70       \\
[3pt]\hline
\end{tabular}\end{center}
\end{small}
\label{table1}
\caption{The  2-d Potts model: Values of $C_o$
from the large $q$ expansion,
exact values for $E_d-E_o$, $C_d-C_o$, and correlation
length $\xi(\beta \to \beta_t^{-})$ of the disordered phase.
The correlation length of the ordered phase is smaller.}
\end{table}

\begin{table} [htp]
\begin{small}
\begin{center}\begin{tabular}{|lll|}\hline
q   &$C_o$      & reference     \\
[3pt]\hline
 7  & 47.5 (25) & \cite{JBK}         \\
    & 50.  (10) & \cite{Paper_q10}   \\
    & 44.  (22) & \cite{Rumm}        \\
[3pt]\hline
10  & 10.6 (11) & \cite{Lee_Kosterlitz_one}    \\
    & 12.7 (3)  & \cite{Paper_q10}      \\
    & $\sim$ 18 & \cite{Paper_q10}      \\
    & 18.0 (2)  & \cite{Janke_Kappler}  \\
[3pt]\hline
15 & 8.04  (4)  & \cite{Janke_Kappler}  \\
[3pt]\hline
20 & 5.2  ( 2)  & \cite{Paper_q20}    \\
   & 5.38 ( 4)  & \cite{Janke_Kappler}  \\
[3pt]\hline
\end{tabular}\end{center}
\end{small}
\label{table4}
\caption{Numerical estimates of the specific heat of
the ordered phase of the $q$ states Potts model.
Results are given in chronological order for each $q$ value.}
\end{table}

Comparison \cite{unpublished} of the predictions of
Eq.\ref{max} and \ref{betat} with numerical
data from lattices of size  $16\leq L\leq 100$ can be found in
Fig.\ref{Fig.CV.max}
and Fig.\ref{Fig.CV.beta} for the  2-d $q=10$ Potts model. In both
figures
the value of the specific heat $CV$ divided by $1/L^d$ is plotted as
function of $1/L^d$. The value at $L=\infty$ is exactly known and
the slope is related to the value of $C_o$ and $C_d$.
An estimate using the large $q$
expansion  of the model is\cite{Tanmoy} $C_o = 18.06 (4)$.
(The difference $C_o-C_d$ is exactly known).
The numerical data for  $CV(\beta_t)$ would allow to
estimate correctly the value for $C_o$. Using  the data
for  $CV_{max}$  or $BL_{min}$ would however led to
underestimate
$C_o$ \cite{unpublished,Paper_q10}.
For a very strong transition, like the one of the $q=20$ model,
results (from similar sized lattices) are unambiguous
\cite{Paper_q20}:  both $CV(\beta_t)$, $CV_{max}$ and $BL_{min}$
give
estimates for $C_o$ that agree with the large $q$ result. On the
other hand
results for the $q=7$ model on lattices as large as $L=64$ do not
enter the asymptotic regime\cite{Paper_q10}.

Published results for $C_o$ from such finite size analysis  can be
found in Tab.\ref{table4} for the $q=20$, $10$ and
$7$ Potts models. The large $q$ expansion estimates can be found
in Tab.\ref{table1}.

\section{DETERMINING THE ORDER OF A PHASE TRANSITION}

It is clear from the above that  equations like Eq.\ref{max} are only
valid in the large volume limit $L>>\xi$. Marginal cases (the so
called weak first order phase transitions) where $\xi$ is very large
and the latent heat is very small are quite hard to discriminate
from second order phase transitions, using numerical methods.

One signature for first order phase transition is the "two peak
signal", namely

i)   $P_L(E)$  has two peaks (of heights  $P^{max}_o$ and
$P^{max}_d$) with a minimum between them of height $P^{min}$.

ii)  This structure behaves the way it should as $L$ grows, namely
the distance between the two peaks goes to a constant
and e.g. $P^{max}_o/P^{min} \to \infty$.

The existence of two peaks in itself is not a signal of a first order
phase
transition. It is well known that, for example, the spin probability
density of the Ising model  has two peaks at the
(infinite volume) transition temperature\cite{Tomaschitz}. The
distance between
the peaks goes to zero as $L$ grows as $L^{-\beta/\nu}$
(here $\beta$ is a critical exponent)
and the ratio of the peak height
to the  minimum height stays constant.

The two peak signal is nearly as old as numerical simulations, it
was made
more precise in Ref.\cite{Lee_Kosterlitz_two}:  Define
$\Delta F = ln (P^{max}_o/P^{min})$, measured at the transition
point
(at zero field for field driven transitions or at the
transition temperature for temperature driven transition.).
If $\Delta F$ increases as $L$
grows the transition is first order, otherwise it is second order.

In the first order case   $\Delta F$ behaves as $L^{d-1}$
for the following reason\cite{Binder_cumulant,Shlosman}: Two phase
configurations with almost planar interfaces dominate in the
region around the minimum of $P_L(E)$. Let $E=xE_o+(1-x)E_d,
x\in [0,1]$, the contribution of such configurations
to the partition function is

\begin{eqnarray}
Z^{\rm mixed}(x) &\o& e^{- x L^d \beta f_o(\beta)}
e^{ - (1-x)L^d \beta f_d(\beta)} \\
\nonumber &\times& L^p e^{-2 \sigma_{o,d}L^{d-1}}  .
\label{Zstrip}
\end{eqnarray}

The first two factors are contributions from the pure phases. The
factor
$L^p \exp (-2 \sigma_{o,d} L^{d-1})$ comes from interface effects,
$\sigma_{o,d}$ is the order-disorder interface tension,
and the exponent $p$ can be computed in a continuous
approximation \cite{CWA}). Finally
$\Delta F \sim  2 \sigma_{o,d} L^{d-1}$
However in current simulations this behavior is only seen for
extremely strong first order phase
transitions, like the $q=20$ Potts model. To
summarize, $\Delta F$ growing with $L$ is
a signal for first order phase transition,
$\Delta F \propto L^{d-1}$ is only seen for
obviously first order cases.

It is a widespread belief that the behavior
of $BL_{min}$  is a better signal for first
order phase transitions that the behavior of
$P_L(E)$ itself. $BL_{min}$ has a non zero
limit for first order transitions, and a zero
limit for second order phase transitions or
out of the precise transition point. Energy and
spin fluctuations have similar
properties. I would like to warn against this
belief:

i) Suppose that, in the simulated range
of lattice size, $P_L(E)$ is approximately
$L$ independent, $BL_{min}$ is then a non
zero constant. Would any one conclude that
the transition is first order? The only sensible
conclusion  is rather that one does not
known the nature of the transition, and
needs simulations on larger systems, in
order to see wether $P_L(E)$  will shrink
like for a second order transition, or  will
develop a two peak structure. Note that there do are
examples in the literature of the use of $BL_{min}$ to show that a
transition is first order in cases where $P_L(E)$  is singled
peaked.

ii) An example of the use of
$BL_{min}$  to determine the order of the
2d $q\in [3,5]$ Potts model phase
transitions can be found in Ref.\cite{Ukawa}
on lattices as large as $L=256$. $BL_{min}$
does seem to go to a non zero limiting value
for $q=5$  and to a zero
limiting value for $q=3$ and $4$. The exact limiting value
for $q=5$ is $BL_{min} = .446 \ 10^{-3}$ (called
$V_{L,min}-{2\over 3}$
in Fig.4 of Ref.\cite{Ukawa}). Looking at the data, it means that
there
must be a crossover for $L>256$. The success of this method should
thus be considered as somewhat accidental.

\section{INTERFACE TENSION}

\begin{figure} [htbp]
\vfill\penalty -5000\vglue 6cm
\includegraphics{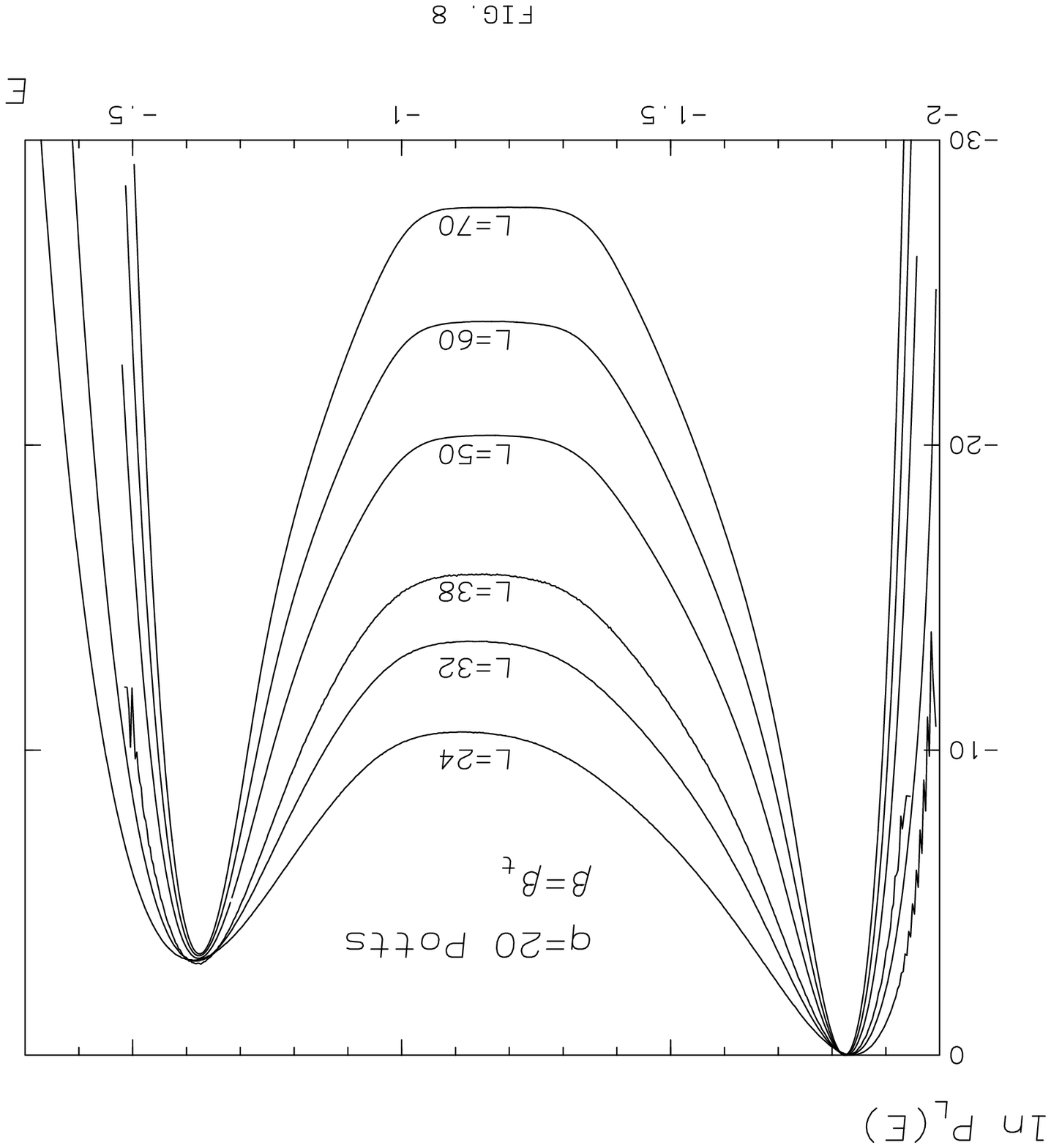}
\caption{ Plot of $\ln~P_L(\beta_t,E)$ for selected lattice sizes in
the q=20
Potts model. The unfolding of a flat region with increasing
lattice size $L\gt 38$ is observed.}
\label{distri}
\end{figure}

The best known technique to determine the order-disorder
interface tension is the histogram method first introduced in
\cite{Binder_Surface}. It uses the two peak structure of
$P_L(E)$, $\sigma_{o,d}$ is estimated as

\begin{equation}
2 \sigma_{o,d} = {1\over 2 L^{d-1}} \ln
\lbrack{P_L^{max,o}P_L^{max,d}
\over (P_L^{min})^2}\rbrack \mid _{\beta=\beta_t}
\end{equation}

with ${\cal O} (1/L^{d-1})$ corrections.
This method became a winner with the introduction of the
multicanonical algorithm \cite{multicanonical}. This algorithm
brings two vast
improvements:
 i) The exponentially large slowing down of the Metropolis
algorithm is strongly reduced, presumably down to a power law
slowing down.
ii) The regions close to the peaks and the region close to the
minimum are sampled equally. Fig.\ref{distri} shows our
results\cite{BBN_Surface} for
the $q=20$ model energy probability distribution. One sees the
unfolding of a flat region between the two peaks, in accordance
with
expectations (Eq.\ref{Zstrip} in $x$ independent
at $\beta=\beta_t$). This unfolding has not (yet) been seen for
lower values
of $q$. Results for $\sigma_{o,d}$ obtained with this method are
compared in Tab.\ref{table2} with the exact result \cite{BW} for
various
values of $q$. Also shown is an estimate using measurements of the
auto-correlation time of the Swendsen-Wang algorithm
\cite{Sourendu}, which behaves like
$\tau \sim L^{d\over 2} e^{2\sigma_{o,d}L^{d-1}}$ \cite{Zinn}. This
estimate is in qualitative agreement with the exact result. The
following two comments are in order

i) It is somewhat surprising to
obtain excellent agreement with the exact results for $q=10$ and
$q=7$, since in both cases $P_L(E)$ does not unfolds its asymptotic
shape.

ii) Ingenious methods\cite{Claudio,Kajantie} where two interfaces
are created by setting one half of the system at a temperature
$\beta_t+\delta\beta_t$ and the other half at  a temperature
$\beta_t-\delta\beta_t$ give results in strong disagreement
with the exact result.

\begin{table} [htbp]
\begin{small}
\begin{center}\begin{tabular}{|l|ll|}\hline
q   & $2 \sigma_{o,d}$ & Reference      \\
[3pt]\hline
 7  & .1886 (12)  & \cite{Claudio}     \\
    & .20         & \cite{Kajantie}    \\
    & .0241  (10)      & \cite{JBK}         \\
    & .020792     & exact              \\
10  & .094701     & exact              \\
    & .0978  (8)  & \cite{multicanonical}     \\
    &  .0950 (5)  & \cite{BBN_Surface}        \\
15  & .239234     & exact               \\
    & .263 (9)    & \cite{Sourendu}      \\
20  & .370988     & exact               \\
    & .3714 (13)  & \cite{BBN_Surface}  \\
[3pt]\hline
\end{tabular}\end{center}
\end{small}
\label{table2}
\caption{Comparison of the numerical estimates of the
order disorder interface tension of the 2-d Potts model
with exact results.}
\end{table}

\begin{figure} [htbp]
\vfill\penalty -5000\vglue 6cm
\includegraphics{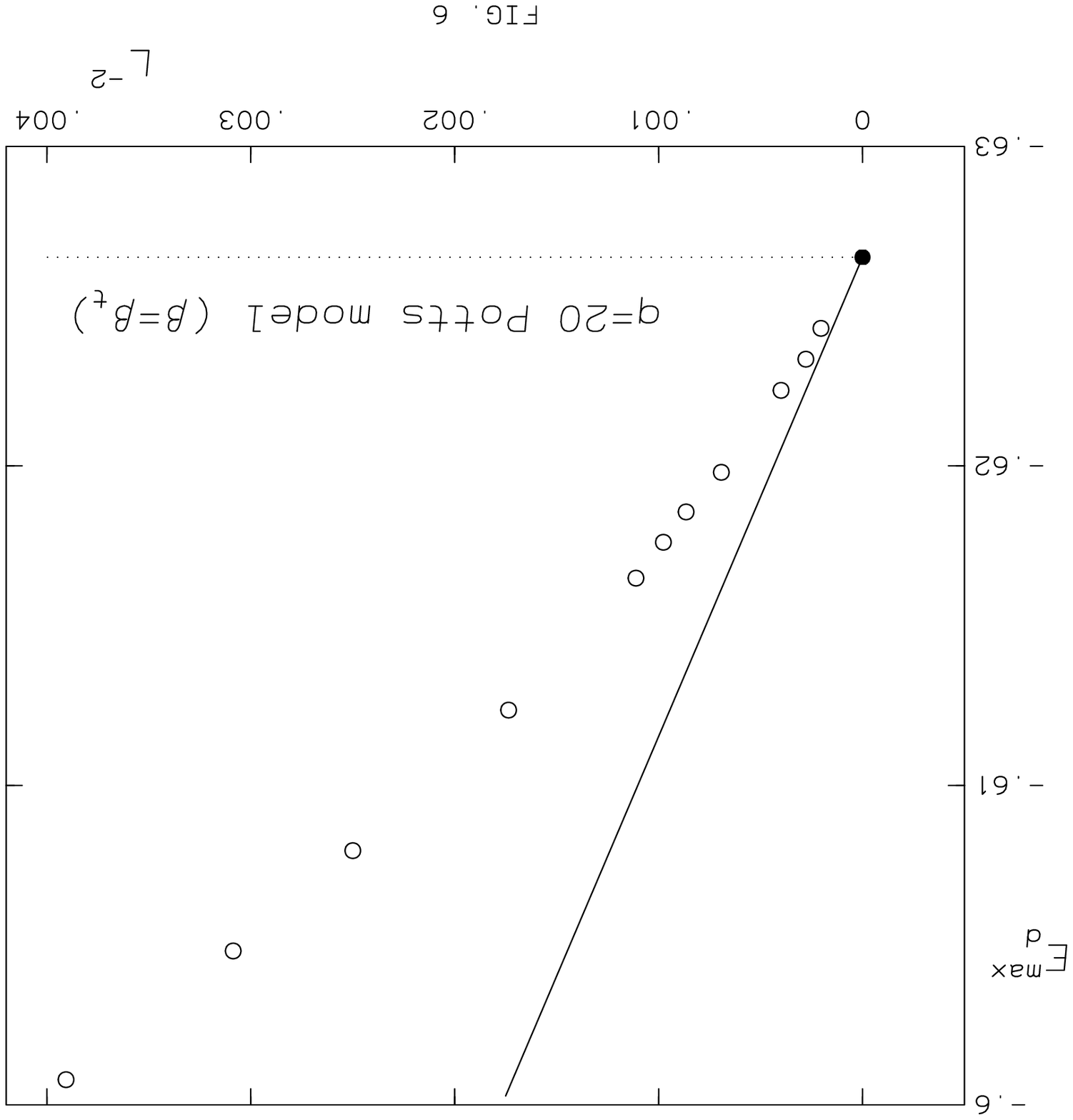}
\caption{   Locations of the maxima of the disordered peak
$E_L^{max,d}$
at $\beta_t$ in the q=20 Potts model as function of the inverse
volume.}
\label{Fig6}

\end{figure}

Among the by products of interface tension measurements are
precise determinations of the locations of the peaks of $P_L(E)$.
These are seen to move with $L$, an effect not described by
Eq.\ref{P(E)}.
Two models have been proposed
to explain this dependence. The first one\cite{Lee_Kosterlitz_one}
involves mixed phase effects, the prediction is $E^{max}_{o,d}(L)
\sim E^{max}_{o,d}(\infty) + A/L$. The second one involves higher
order $1/L^d$ terms in the Laplace transform of Eq.\ref{Z},
neglecting all ${\cal O} (e^{-b L})$ terms, namely all mixed phase
effects. The  result is $E^{max}_{o,d}(L) \sim E^{max}_{o,d}(\infty) +
B/L^d$, where $B$ is related to the third derivative of the free
energy of the phase. Numerical estimate \cite{BBN_Surface} of
$E^{max}_{d}(L) $ as function of $1/L^d$ are shown in Fig.\ref{Fig6}
for
the $q=20$ model. The line drawn is the parameter free prediction
of model  two from Ref.\cite{Tanmoy} in leading order.
Similar results are obtained for the $q=10$ case. The predicted
$1/L^2$ behavior is barely seen, but i) It has been shown in
Ref.\cite{Tanmoy} that the discrepancy can be accommodated by
higher orders terms (still without any mixed phase effects) ii) A fit
to
the $1/L$ form gives an estimates of  $E^{max}_{d}(\infty) $
incompatible with the exact value of $E_d$.

\end{document}